# INVESTIGATING GALAXY EVOLUTION WITH *CHANDRA*

G. Fabbiano

*Harvard-Smithsonian Center for Astrophysics, 60 Garden St., Cambridge MA 02138, USA*

**ABSTRACT**

*Chandra* observations show the importance of the X-ray band for studying the evolution of galaxies. Binary X-ray sources are an easily detectable tracer of the stellar population. *Chandra* studies of these populations are giving us insights into the nature and formation of these binaries, and provide the basis for diagnostics of galaxy evolution. With *Chandra* and *XMM-Newton* we can explore relatively poorly known aspects of the black hole population of the universe: ultra-luminous X-ray sources, that may be connected with the 'missing' intermediate mass black holes predicted by hierarchical galaxy and black hole formation scenarios; and quiescent supermassive nuclear black holes and their surroundings, as a way of understanding the full range of the AGN phenomenon. Finally, the X-ray band provides the only way to explore hot plasmas in galaxies; recent observations are revealing the importance of these plasmas as vehicles of both chemical enrichment and energy.

## 1. INTRODUCTION

This talk would not have been possible without *Chandra*. Although *XMM-Newton* has contributed significantly to the study of the nearest galaxies, the sub-arcsecond resolution of *Chandra* is essential for detecting populations of X-ray sources in galaxies to the Virgo Cluster and beyond, at the luminosities of the bright Galactic X-ray binaries.

This resolution is also needed to explore the X-ray emission of normal galaxies at high redshift, to obtain sensitive data on the emission and the surroundings of the silent supermassive black holes found at the nuclei of most large galactic bulges, and to study the relatively uncontaminated emission of hot plasmas in galaxies.

## 2. STELLAR EVOLUTION IN X-RAY BINARIES

It is well known that the Milky Way hosts both old and young X-ray source populations, reflecting its general stellar make up. With *Chandra*'s sub-arcsecond angular resolution, combined with CCD photometric capabilities (Weisskopf *et al.* 2000), we can now study these X-ray populations in galaxies of all morphological types, down to typical limiting luminosities in the $10^{37}$ ergs s$^{-1}$ range. At these luminosities, the old population X-ray sources are accreting neutron star or black-hole binaries with a low-mass stellar companion, the LMXBs (life-times $\sim 10^{8-9}$ yrs). The young population X-ray sources, in the same luminosity range, are dominated by neutron star or black hole binaries with a massive stellar companion, the HMXBs (life-times $\sim 10^{6-7}$ yrs; see Verbunt & van den Heuvel 1995 for a review on the formation and evolution of X-ray binaries), although a few young supernova remnants (SNRs) may also be expected. At lower luminosities, reachable with *Chandra* in Local Group galaxies, Galactic sources include accreting white dwarfs and more evolved SNRs. With *Chandra*'s angular and spectral resolution, populations of point-like sources are easily detected above a generally cooler diffuse emission from the hot interstellar medium (fig. 1). Note that luminous X-ray sources are relatively sparse by comparison with the underlying stellar population, and can be detected individually with the *Chandra* sub-arcsecond resolution, with the exception of those in crowded circum-nuclear regions.

To analyze this wealth of data two principal approaches have been taken: (1) a photometric approach, consisting of X-ray color-color diagrams and color-luminosity diagrams, and (2) X-ray luminosity functions (XLFs). Whenever the data allow it, time and spectral variability studies have also been pursued. Optical and radio identifications of X-ray sources and association of their position with different galaxian components are now being increasingly undertaken.

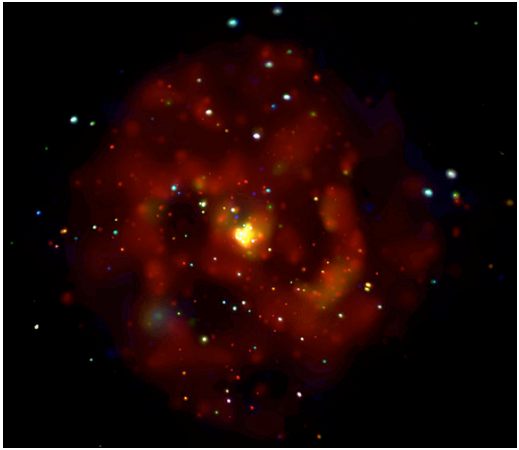

Fig. 1 – *Chandra* image of M83 (red is low energy and blue high energy), from the *Chandra* web page (Soria & Wu 2003)

Given the lack of standard X-ray photometry to date, different definitions of X-ray colors have been used in different works; in the absence of instrument corrections, these colors can only be used for comparing data obtained with the same observational set up. Colors, however, have the advantage of providing a spectral classification tool when a limited number of photons are detected from a given source, which is certainly the case for most X-ray population studies in galaxies. Also, compared with the traditional derivation of spectral parameters via model fitting, color-color diagrams provide a relatively assumption-free comparison tool. *Chandra*-based examples of this approach can be found in Zezas *et al.* 2002a, b and Prestwich *et al.* 2003, among others. The X-ray color-color diagram of Prestwich *et al.* 2003 (fig. 2) illustrates how colors offer a way to discriminate among different types of X-ray sources.

XLFs are increasingly used to characterize the X-ray source populations of different galaxies. Compared to the Milky Way, external galaxies provide clean source samples, all at the same distance. Moreover, the detection of X-ray source populations in a wide range of different galaxies allows us to explore global population differences that may be connected with the age and or metallicity of the parent stellar populations (see review of Fabbiano & White 2005 and references therein; Kong *et al.* 2003; Belczynski *et al.* 2004). In general, X-ray sources associated with young stellar populations follow a significantly flatter XLF than that of the X-ray sources in old stellar systems. A good example is provided by M81, where the XLF of the spiral arm stellar population is flatter than that of the inter-arm and bulge regions, consistent with the prevalence of short-lived luminous HMXBs in younger stellar populations (Tennant *et al.* 2001, fig. 3; Swartz *et al.* 2002).

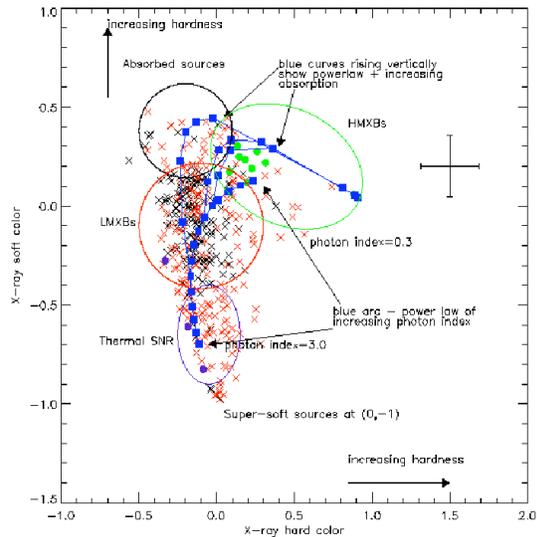

Fig. 2 – *Chandra* color-color diagram from Prestwich *et al.* 2003

The same trend is found comparing the X-ray populations of actively star-forming galaxies with those of E and S0s. While the total number of X-ray sources in star-forming galaxies (the normalization of the XLF) is driven by the star formation rate, in E and S0 galaxies total stellar mass appears to be the driving factor, with the specific frequency of globular clusters as a second order effect (e.g., Zezas & Fabbiano 2002; Kilgard *et al.* 2002; Grimm, Gilfanov & Sunyaev 2003; Gilfanov 2004; Kim & Fabbiano 2004).

The tools that are being developed for characterizing and understanding the X-ray source populations of nearby galaxies lay the foundation of future work in X-ray population synthesis (see Belczynski *et al.* 2004). We know from the co-added statistical detections of high redshift galaxies in deep *Chandra* surveys that there is X-ray evolution with redshift (Lehmer *et al.* 2005) in galaxies. The next step will be to use what we are learning from the nearby universe, together with the observational constraints derived from observations of the high $z$ universe, to put firm observational and theoretical constraints on the evolution of X-ray binary populations (e.g. Ghosh & White 2001).

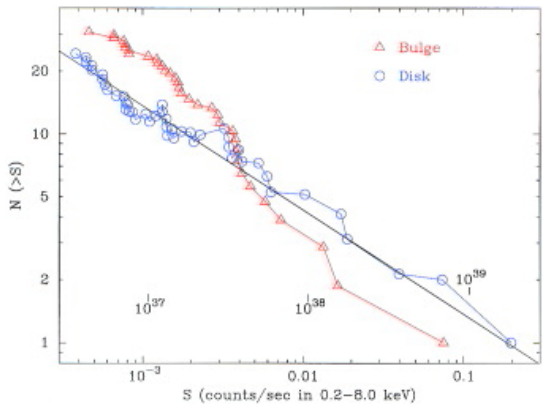

Fig. 3 – Bulge and disk XLFs of M81 (Tennant *et al.* 2001).

## 3. BLACK HOLES AND THEIR ENVIRONMENT

One of the current hot topics in astrophysics and cosmology is the formation and evolution of black holes and their relation to the formation of galaxies. X-ray observations can help constrain some of these scenarios. I will discuss here two topics related to understanding the entire spectrum of black holes: ultra-luminous X-ray sources (ULXs), and silent nuclear supermassive black holes.

### 3.1 ULXs

The most widely used observational definition of ULXs is that of sources detected in the X-ray observing band-pass with luminosities of at least $10^{39}$ erg s$^{-1}$, implying bolometric luminosities clearly in excess of this limit. This ULX luminosity is significantly in excess of the Eddington limit of a neutron star ($\sim 2 \times 10^{38}$ erg s$^{-1}$), suggesting accreting objects with masses of 100 $M_\odot$ or larger. Since these masses exceed those of stellar black holes in binaries (up to $\sim$30 $M_\odot$, Belczynski, Sadowski & Rasio 2003), ULXs could then be a new class of astrophysical objects, possibly unconnected with the evolution of the normal stellar population of a galaxy. ULXs could represent the missing link in the black hole mass distribution, bridging the gap between stellar black holes and the super-massive black holes found in the nuclei of early type galaxies. These 'missing' black holes have been called intermediate mass black holes (IMBH), and could be the remnants of hierarchical merging in the early universe (Madau & Rees 2001), or could be forming in the core collapse of young dense stellar clusters (e.g. Miller & Hamilton 2002).

While ULXs have been known for the last ~20 years, the detection of large samples of these sources has required the observations of many galaxies, and in particular active star-forming systems, where they are copious. *Chandra* and *XMM-Newton* observations have shown that ULXs tend to be associated with very young stellar populations (such as that of the Antennae galaxies, where 14 such sources are found; Fabbiano *et al.* 2004a). These results have suggested the alternative view that ULXs could just represent a particular high-accretion stage of X-ray binaries, possibly with a stellar black hole accretor (King *et al.* 2001; see also Grimm, Gilfanov & Sunyaev 2003; Rappaport, Podsiadlowski & Pfahl 2005), or even be powered by relativistic jets in microquasars (Koerding, Falke & Markoff 2002). *Chandra* and *XMM-Newton* work has confirmed that ULXs are compact accreting sources (as suggested by *ASCA* results, Makishima *et al.* 2000; Kubota *et al.* 2001). Flux-color transitions have been observed in a number of ULXs, suggesting the presence of an accretion disk. Some of these spectra and colors are consistent with or reminiscent of those of black hole binaries (e.g., in the Antennae, Fabbiano *et al.* 2003a, b, 2004a). However, these results do not constrain unequivocally the mass of the compact accretor (see e.g. reviews by Fabbiano & White 2005; Fabbiano 2005a).

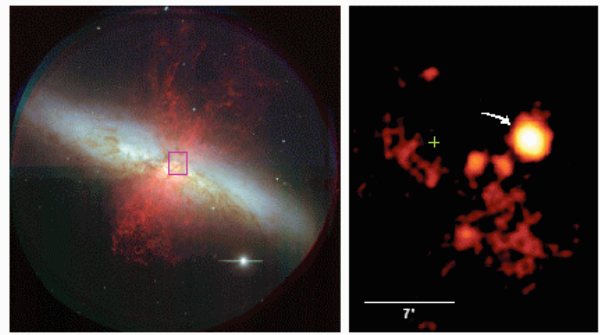

Fig. 4 - M82, left, and the *Chandra* image of the area in the marked rectangle (right), with the ULX marked by the arrow; from Fabbiano (2005b)

My present view (based on the observational results to date) is that ULXs may be a mixed bag of sources, perhaps a few are the elusive IMBHs, but most may be more normal black hole binaries (see Fabbiano 2005b).

### 3.2 Silent supermassive black holes

It is now established that supermassive nuclear black holes are ubiquitous in large E and S0 galaxies and in spiral bulges (e.g. Tremaine *et al.* 2002). Only a small fraction of these black holes are luminous AGN, while low-level activity is more widespread. Some of these nuclei are 'silent' (Ho, Filippenko & Sargent 2003). With *Chandra* we can now explore the emission properties of these low-activity and silent supermassive black holes down to luminosities typical of X-ray binary emission; we can also set constraints on the hot fuel available for accretion and examine the surrounding hot interstellar medium for evidence of past outbursts of activity (Fabbiano *et al.* 2003, 2004b, Jones *et al.* 2002). Fig. 5 shows four of these silent nuclei (Soria *et al.* 2005). All these early-type galaxies by selection have measured nuclear masses, and stringent upper limits on optical line and radio emission. The sources we detect with *Chandra* have luminosities in the range $10^{38}$-$10^{39}$ erg s$^{-1}$ and tend to be associated with extended emission, from which an estimate of the Bondi accretion parameters can be derived. In these nuclei (see also Fabbiano *et al.* 2004b; Pellegrini 2005), accretion must be inefficient, but ADAF accretion, fueled by both the hot ISM and by stellar out-gassing, can explain the emission. Cyclic activity and outflow cycles (e.g., Binney & Tabor 1995; Ciotti & Ostriker 2001) may avoid accumulating large amounts of material in the immediate surroundings of the black holes.

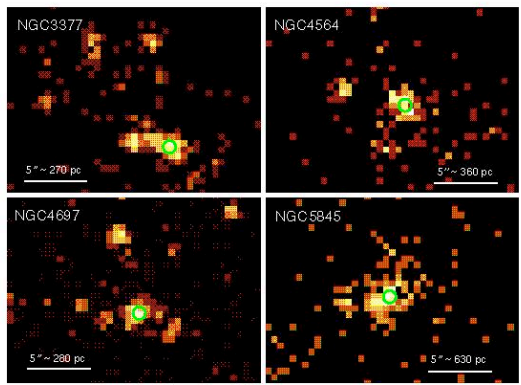

Fig. 5 – Four examples of *Chandra* images of silent supermassive black holes (Soria *et al.* 2005)

Why does a black hole awaken? It is not clear. Galaxy interaction and merging have been suggested in the past as triggers of activity, but there isn't yet a strong statistical evidence of this effect. Recent X-ray observations have provided evidence of nuclear activity in strongly interacting galaxies, suggesting that at least in some cases interaction does indeed facilitate nuclear accretion. NGC6240 is a particularly impressive example (Komossa *et al.* 2003). In this late merger galaxy two X-ray hard nuclei have been found with *Chandra* ACIS, and their spectra show clear Fe K emission lines.

If AGN activity is intermittent, perhaps resulting from a feedback cycle (e.g., Ciotti & Ostriker 2001), some evidence of past outbursts may be present in the hot ISM. The spiral-like feature in NGC4636 could be such a remnant (Jones *et al.* 2002), and so perhaps could be a faint elongated structure in NGC821 (Fabbiano *et al.* 2004b). In NGC5128 (Cen A), a huge gaseous ring perpendicular to the jet was discovered with *Chandra*. Its properties are consistent with it being shock heated gas from a past nuclear outburst, possibly connected with the recent merger event in Can A (Karovska *et al.* 2002; fig. 6).

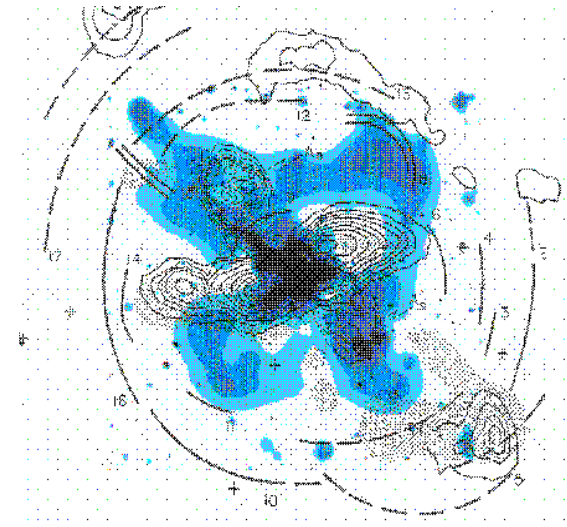

Fig. 6 – In blue is the X-ray emission of NGC5128 detected with *Chandra*. Note the jet and the 'hot ring' surrounding the nucleus (from Karovska *et al.* 2002).

## 4. HOT WINDS AND THE ECOLOGY OF THE UNIVERSE

Starting with the first Einstein observations (see Fabbiano 1989) it was clear that hot gas and galactic winds are present in actively star-forming galaxies, such as M82 and NC253 in the nearby universe. Through these winds the star-forming galaxies will influence their environment, increasing its entropy, and also depositing newly formed elements into the intergalactic medium. Understanding these winds is therefore important if we want to fully understand the ecology of the universe.

With *Chandra* and *XMM-Newton* we can now get a significantly deeper understanding of these hot gaseous components. I will concentrate here on a recent example, resulting from the deep *Chandra* observations of the Antennae galaxies (NGC4038/39), the prototypical galaxy merger. This system was observed with *Chandra* ACIS for 411 ks. resulting in a spectacular data set (fig. 7, Fabbiano *et al.* 2004a). The hot ISM of the Antennae is discussed in detail by Baldi *et al.* (2005a, b; see also this conference), so I will not talk about it here. I will instead discuss the still mysterious large-scale features seen in this hot ISM, extending well beyond the optical bodies of the merging galaxies. These giant loops extend for ~10kpc to the south of the Antennae, and are embedded in lower surface brightness diffuse emission, that can be traced out to 20 kpc. The loop temperature, from the *Chandra* data, is ~0.3 +/- 0.02 keV, significantly larger that the that of the more diffuse halo (0.23 +/- 0.02 keV), possibly suggesting adiabatic cooling and an expanding halo or wind. We do not know what causes these loops, nor what their future evolution will be. Two possibilities are starburst-blown bubbles or merger-induced shocks. From an energy budget point of view, there is plenty of supernova energy deposited in the starburst to cause superbubbles. Given the parameters of the merger, the loops could be propagating with a velocity of 100-1000 km s$^{-1}$, to be compared with a sound speed of ~200 km s$^{-1}$. So it is possible that the loops are due to shock-heated gas, but good spectral data are missing. The *Chandra* ACIS data do not have the necessary signal to noise ratio to discriminate among different options. In particular, if the loops are superbubbles, one would expect a cooler interior (Castor, McCray & Weaver 1975). If they are outwardly propagating shocks, the outer rim should be cooler. A deep *XMM-Newton* observation would answer these questions.

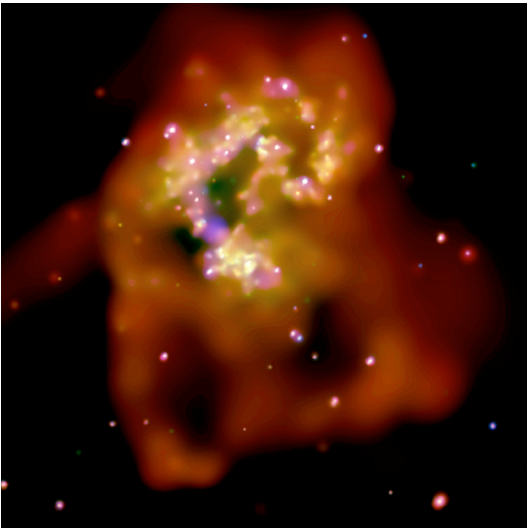

Fig. 7 – Deep *Chandra* image of the Antennae (NGC4038/39), Fabbiano *et al.* (2004a)

The loops could be the result of merger interactions. With accurate temperature and density maps one could attempt a comparison with model simulations (e.g., Barnes 2004). The *Chandra* data do not provide enough statistics for a detailed spectral mapping of these features, but a deep *XMM-Newton* observation could.

Deep *XMM-Newton* data could also address the physical status of the diffuse halo (in equilibrium or expanding), and its metal content. Since we have learned from the deep *Chandra* observation that the hot ISM of the Antennae is metal enriched by SNII events (see Baldi *et al.* 2005a, b), it would be important to measure the metal content of the halo, and compare it with the metal production in the starburst, because this would provide prima facie evidence of how the transport of metals in the intergalactic medium may occur. Obtaining a spectral map of the halo, which would be possible with *XMM-Newton*, would also address crucial questions for the physical status and evolution of the halo. In particular, is the central entropy raised as in galaxy groups (Lloyd-Davies, Ponman & Cannon 2000)? Is the profile at large radii steep and cooler, suggesting winds? We expect that the Antennae will eventually evolve into an elliptical galaxy. Since the mass in the halo is comparable to the amount of diffuse hot gas detected in X-ray faint E and S0 galaxies, and the cooling time is long (Fabbiano 2004a), this type of deep data would provide unique constraints on models of halo development in elliptical galaxies.

Concluding, X-ray observations have discovered a hot gaseous component in galaxies and are now beginning to reveal the physical status and chemical composition of this component. This hot component needs to be included in any simulation of galaxy and merging evolution.

## 5. PROSPECTS FOR INVESTIGATING GALAXY AND BLACK HOLE FORMATION

We are now witnessing a revolution in the study of galaxies in X-rays. We have progressed from the discovery and characterization phase to using the X-ray window as an important part of our understanding of the evolution of galaxies. Given the connection between HMXB populations and star formation in galaxies, illustrated by X-ray population studies of galaxies with *Chandra*, deep X-ray observations give us a means to measure directly the star-formation rate in the deep universe (e.g., Ranalli, Comastri & Setti 2003; Grimm, Gilfanov & Sunyaev 2003). With *Chandra*'s high-resolution telescope we can directly study the interaction and feedback between nuclear black holes and the host galaxies, an important ingredient in present day cosmological simulations (e.g., Granato *et al.* 2004; Okamoto *et al.* 2005). Moreover, X-ray observations of the hot gaseous component of galaxies have demonstrated that gravity is not the only important force in galaxy formation and evolution. Mergers shock-heat the ISM/IGM and increase its entropy. SN, active nuclei, and perhaps dark jets in X-ray sources all alter the energy budget by heating the ISM, and producing galactic winds. Stellar evolution enriches these winds with chemical elements, and therefore alters the chemistry of the universe at large. X-ray observations provide a direct observational window into these phenomena.

Although *XMM-Newton* has and will contribute substantially to this progress, the sub-arcsecond resolution of *Chandra* has been the true catalyst of this revolution. It is *Chandra*'s resolution, and the resulting sensitivity, that has allowed the detections of samples of X-ray sources d'wn to Galactic XRB luminosities in galaxies more distant than the Virgo Cluster; these samples of X-ray sources have permitted X-ray population studies, providing a probe of the evolution of X-ray binaries in a variety of different environments, and have led to the detection of extreme sources, such as ULXs, in copious numbers. It is *Chandra*'s resolution, and the associated spectral capabilities, that have allowed the separation of point like sources and hot diffuse emission in galaxies, leading to the discovery of metal enrichment in these gases. Finally, it is this resolution that has made possible the study of the faintest reaches of nuclear activity.

While *Chandra* is beautiful, it is a small telescope, and this means that forbiddingly long exposure times are needed to exploit some of these results to the full. For example, a dedicated week of exposure time was needed to obtain the beautiful data on the Antennae that I have shown in this talk. Evolution is only inferred by stacking data on the positions of Hubble deep image galaxies, because individual high redshift galaxies cannot be detected with *Chandra* (Lehmer *et al.* 2005). There is, however, recognition of the potential of X-ray astronomy for studies of cosmology and galaxy and black hole evolution. This recognition has resulted in the *Generation-X* proposal, approved for study by NASA, for a very large future X-ray telescope (~100 square meters mirror), with 0.1 arcsecond resolution. This telescope will not be deployed before 2025.

But this is far away in the future. As a community, we should make sure that this resolution is not lost once *Chandra* stops operating, and that future larger telescopes match or even surpass it. Unfortunately, there are no planned missions (either by NASA or by ESA) that will carry the legacy of *Chandra* forward in the near or foreseeable future. I see this as a serious problem not only for X-ray astronomy, but also for astronomy as a whole.

This work was partially supported under NASA contract NAS8-39073 (CXC).
Some material was also covered in a review talk delivered at the COSPAR Colloquium "Spectra and Timing of Compact X-ray Binaries", held in Mumbai (India), January 17-20, 2005, and in a review talk delivered at the IAU Symposium 230 "Populations of High Energy Sources in Galaxies" held in Dublin (Ireland), August 15-19, 2005.